# Spin-filter modeling by means of extension theory methods

**L A Dmitrieva[1], D N Krupa[2] and Yu A Kuperin[3]**

[1] Division of Mathematical Physics, Department of Physics, Saint Petersburg State University, 198504 Saint Petersburg, Russia;

[2] Division of Computational Physics, Department of Physics, Saint Petersburg State University, 198504 Saint Petersburg, Russia

[3] Division of Nuclear Physics, Department of Physics, Saint Petersburg State University, 198504 Saint Petersburg, Russia

E-mail: **ludmila.dmitrieva@gmail.com** (Dmitrieva), dimakrupa@gmail.com (Krupa), **yuri.kuperin@gmail.com** (Kuperin)

**Abstract**

The problem of spin-dependent transport of electrons through a finite array of quantum dots attached to 1D quantum wire (spin gun) for various semiconductor materials is studied. Unlike the model considered in [1] a model proposed here is based on the extension theory model (ETM) and assumes the quantum dots to have an arbitrary internal structure, i.e. the internal energy levels. The presence of internal structure in quantum dots results in energy-dependent interaction between electrons and quantum dots. This interaction changes the transmission mode of the spin current through the spin gun. For the energy-dependent interaction it is shown in this article the difference of transmission probabilities for singlet and triplet channels for several quantum dots in the array due to interference effects can reach approximately 100% percent for some energy intervals. For the same energy intervals the conductance of the device reaches the value $\approx 1$ in $[e^2/\pi\hbar]$ units. As a result a model of the spin-gun which transforms the spin unpolarized electron beam into completely polarized one is suggested.

PACS numbers: 02.30.Tb, 03.65.Nk, 05.60.Gg

**1. Introduction**

Spintronics is a rapidly developing branch of nanoscience which exploits spin properties of electrons or nuclei instead of charge degrees of freedom [2]. Perhaps the most current efforts in designing and manufacturing the spintronics devices is focused now on finding novel ways of both generation and utilization of spin-polarized currents. This, in particular, includes the study of spin-dependent transport in semiconductor nanostructures, which can operate as spin polarizer or spin filters. Although spintronics can have a lot of practical applications [3] - [11], one of them is most interesting and ambitious: the application of electron or nuclear spins to quantum information processing and quantum computation [12], [13].

Unlike the model of spin filter proposed in [1] in this paper quantum dots that make up the filter endowed with an internal structure. To simulate the interaction of internal degrees of freedom of quantum dots and the external degrees of freedom of electrons in a quantum wire, we use the extensions theory of symmetric operators (see e.g. [14, 15]). The model of spin filter proposed here is more realistic than the model built in [1]. In addition, as shown below an energy-dependent generalization of the Breit-Fermi interaction has been constructed. As a result we suggest a new type of spin polarizer constructed from the finite array of semiconducting quantum dots (QD) attached to a quantum wire (QW). All dots of the array have an internal structure, are identical and each of them carries the spin 1/2. The schematic construction of the proposed device is shown for a single dot in figure 1 and for the finite array of dots in figure 2.

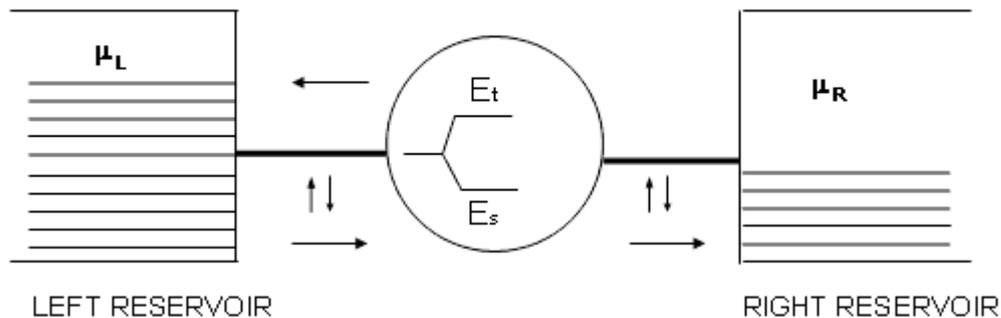

**Figure 1.** Schematic representation of one quantum dot attached to quantum wires. The left and right reservoirs are at different chemical potentials $\mu_L$ and $\mu_R$. The horizontal arrows represent incident, transmitted and reflected electronic wave. The quantum dot carries spin s=1/2. By $E_s$ and $E_t$ the singlet and triplet states are denoted. By $\uparrow\downarrow$ the electron spin projection before and after the scattering by the quantum dot is denoted. If the small voltage $eV = \mu_L - \mu_R$ is applied a non-equilibrium situation is induced and spin-dependent current will flow through the single dot device.

We consider the following model of quantum dots. Using the extension theory model (ETM) we endow each quantum dot with arbitrary internal structure (see, e.g. [14], [15]). The parameters of internal structure are the energy levels of quantum dot and matrix elements of corresponding spectral projections. These parameters should be fitted in solving the scattering problem for the spin filter device with taking into account the properties of specific semiconductor material. Let us note that for one electron scattering or the scattering of electron beam with interaction between electrons being neglected this problem is exactly solvable in frames of ETM.

For quantum dots having internal structure in the first place we give mathematical construction (see Section 3 and 4). After that in terms of scattering data, in particular, in terms of spin-dependent transmission coefficients through the device, we calculate the most important measurable characteristics: the polarization efficiency $P(k)$ and the conductance $G(k)$ of the device (see Section 5). By explicit calculations we show that there exist some intervals of "admissible" momentum $k$ of incident electrons for which the polarization efficiency $P(k) \approx 100\%$ and at the same intervals the conductance $G(k)$ approaches a half from its upper limit $e^2/\pi\hbar$. It means that at such values of electrons momentum the unpolarized beam of incoming electrons transforms into fully polarized outgoing electrons beam and simultaneously at the same values of momentum the device possesses a high conductance. Because of these properties we have called the proposed device as spin gun. The pilot results on spin gun construction in case of quantum dots having internal structure can be found in [16].

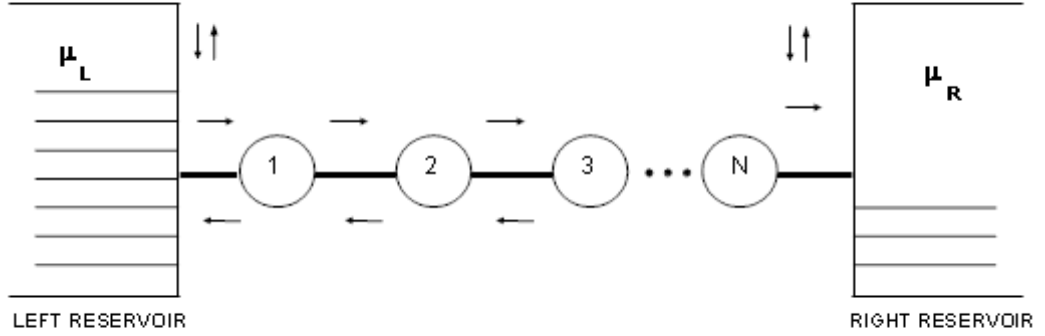

**Figure 2.** Schematic representation of spin filter device (spin gun) fabricated from N quantum dots and quantum wires. For details see the caption for figure 1.

**2. Assumptions, constrains and choice of model parameters**

In this section we formulate the physical conditions under which our mathematical modeling is relevant. We also discuss the physical and technological constrains connected with a choice of semiconductor materials which guarantees that spin gun device can be in principle manufactured. Finally we discuss the fit of the model parameters and possible range of their variation which does not destroy the operational regime of the device, i.e. the ability to produce the polarized beam of electrons and simultaneously to conduct spin-dependent current at sufficiently high level.

We assume that in the quantum wire connected with a quantum dots one-mode propagation of electrons in a ballistic regime can be realized. For the ballistic propagation of electrons through the QW the de Broglie wave-length [17] $\lambda_B = 2\pi\hbar(2m^*k_BT)^{-1/2}$ has to be much greater than both the mean free path $l_{mfp}$ in the material of QW and the length $d$ of QW between the neighboring quantum dots, i.e. $\lambda_B \gg l_{mfp}$ and $\lambda_B \gg d$.

On the other hand in order to use zero-range potentials for mathematical modeling of scattering by quantum dots the de Broglie wavelength has to be greater than the mean size $r_0$ of the dot: $\lambda_B \gg r_0$ [18]. There are different ways to manufacture quantum dots (see, e.g. [19], [20], [21]). Depending on technology the mean size of quantum dots varies from 20 nm for small dots up to more than 100 nm for the large dots. In order to satisfy all constrains mentioned above we have to assume that spin-gun can be realized on relatively small QD using narrow-gap semiconductors. For such materials (see table 1) $\lambda_B$ is about 50 nm at room temperature and more than 100 nm at the liquid nitrogen temperature. It means that we have either to use in our modeling the parameters $d, r_0 \ll 100$ nm or to assume the working temperature of the device less than $T = 77$ K. Below in Section 5 we shall take into account these limitations in our numerical simulations of the device operation.

**Table 1.** Physical characteristics of narrow-gap semiconductors [22]

| Semiconductor | $E_g$, eV (285 K) | $\frac{m^*}{m_e}$ | $\lambda_B$, nm (285 K) | $\lambda_B$, nm (77 K) |
|---|---|---|---|---|
| GaAs | 1.430 | 0.068 | 29.9 | 57.7 |
| InAs | 0.360 | 0.022 | 52.8 | 101.5 |
| $Cd_xHg_{1-x}Te$ | | | | |
| $x = 0.20$ | 0.150 | 0.013 | 68.6 | 131.9 |
| $x = 0.30$ | 0.290 | 0.021 | 54.0 | 103.9 |
| $x = 0.44$ | 0.510 | 0.055 | 33.4 | 64.2 |
| HgTe | -0.117 | 0.012 | 71.4 | 137.4 |
| $Zn_{0.15}Hg_{0.85}Te$ | 0.190 | 0.015 | 63.9 | 122.9 |
| InSb | | 0.014 | 66.1 | 127.2 |

## 3. A mathematical model of spin transport through a single quantum dot with internal structure

### 3.1. Preliminaries on extension theory methods

In this section we model the interaction between electrons and the array of quantum dots by means of Extension Theory Methods (ETM). There is a lot of publications devoted to the ETM applications to modeling various phenomena in quantum and classical physics. Here we mention only few references on the subject [14], [15], [23], [24].

The main idea is the following. We have to take a pair of self-adjoint operators, describing different degrees of freedom of the system in question, and acting in different Hilbert spaces. The first operator describes an internal structure of each quantum dot which are assumed to be equivalent. The second operator describes the electron propagation in the quantum wire connecting quantum dots. In the simplest case we choose this operator as kinetic energy of electron. We have to restrict both of them to the symmetric operators having the so-called deficiency subspaces. Then we have to combine them (say, to add them) in such a way that we obtain a symmetric operator, acting in the sum of Hilbert spaces. This symmetric operator has the deficiency subspaces lying in the sum of Hilbert spaces. We have to consider the adjoint operator to the symmetric operator in question. The domain of this adjoint operator is given by the sum of the domain of the symmetric operator and its deficiency subspaces (the first von Neumann theorem). The next step is it to extend the symmetric operator to a self-adjoint one. The universal way to construct this extension is given by the second von Neumann theorem, namely we have to choose in the deficiency subspaces some linear set such that the so-called symplectic boundary form vanishes. If we do all of that then we obtain a new self-adjoint operator, acting in the pair of Hilbert spaces. This operator is treated in the ETM as a Hamiltonian with interaction between internal and external degrees of freedom of the system. This procedure has an abstract character and can be applied for any type of initial self-adjoint Hamiltonians. In our modeling the interaction between internal (quantum dot) and external (electrons in quantum wire) degrees of freedom including the spin degrees of freedom of the both subsystems appears as the difference between the final self-adjoint operator and orthogonal sum of two initial operators mentioned above. After the joint self-adjoint extension of the internal and external Hamiltonians the total operator describing the common dynamics of external and internal degrees of freedom acts in the orthogonal sum of two Hilbert spaces which we shall call the external and internal channels correspondingly. Since the transport process of electrons takes place in the external channel only, the reduction by means of ETM of the system to the external channel leads to energy-dependent interaction in the external channel. In our model such type of exclusion of internal degrees of freedom leads to the similar type of boundary conditions as we obtained in our previous model with Breit-Fermi spin-spin interaction without internal structure of quantum dots [1]. The only difference is that the boundary conditions will be energy-dependent and this dependence is parameterized by the some function of energy with positive imaginary part in the upper complex half-plane. This function is quite standard in frames of ETM and is called the Schwartz integral. As it is shown in frames of ETM this function caries the whole information about the dynamics of the internal degrees of freedom. In what follows we adduce a brief description of the corresponding mathematical constructions.

### 3.2. The choice of the state space and the structure of the Hamiltonian in external channel

Let $\mathcal{H}_s$ be a single spin space $\mathcal{H}_s \simeq \mathbb{C}^2$ and $I_s$ be the identical operator in $\mathcal{H}_s$. As in the previous model [16] we suppose $s = 1/2$ and $s_3 = \pm 1/2$ both for the electrons and quantum dot. Since the external space without spin $\mathcal{H}^{ex}$ for one-dimensional quantum wave-guide is $L^2(\mathbb{R})$ the free Hamiltonian is self-adjoint (s.a.) operator $\mathcal{H}^{ex}$ is given by the expression (in the units $\hbar = 2m^* = 1$)

$$H^{ex} = -\frac{d^2}{dx^2} \qquad (1)$$

on the domain $D(H^{ex}) = W_2^2(\mathbb{R})$, where $W_2^2(\mathbb{R})$ is the Sobolev space which sometimes is denoted as $H^{2,2}(\mathbb{R})$ and $m^*$ is the effective mass of electron in the quantum wire.

After taking into account the spin of electrons the free Hamiltonian in the external channel (space) is obviously s.a. operator $H_s^{ex} = H^{ex} \otimes I_s$ acting on the domain $D_s(H_s^{ex}) = W_2^2(\mathbb{R}) \otimes H_s \simeq W_2^2(\mathbb{R}, \mathbb{C}^2)$. So the elements of $D_s(H^{ex})$ are $\psi^{ex} = \begin{pmatrix} \psi_{+1/2}^{ex} \\ \psi_{-1/2}^{ex} \end{pmatrix}$, where $\psi_\alpha^{ex} \in W_2^2(\mathbb{R})$, $\alpha = \pm 1/2$.

### 3.3. The choice of the state space and the structure of the Hamiltonian in internal channel

Let $\mathcal{H}^{in}$ be an arbitrary finite-dimensional Hilbert space, $\dim \mathcal{H}^{in} = n$ and $H^{in}$ is an arbitrary s.a. operator in $\mathcal{H}^{in}$ with the simple spectrum:

$$H^{in} = \sum_{j=1}^{n} E_j < \cdot, f_j > f_j, \ E_j \neq E_k, j \neq k \qquad (2)$$

where $E_j$ and $f_j$ are eigenvalues and corresponding eigenvectors of the operator $H^{in}$ and $<\cdot,\cdot>$ is the scalar product in the inner space $\mathcal{H}^{in}$.

In order to describe the symmetric restriction $H_0^{in}$ of the s.a. operator $H^{in}$ we use the universal technique of ETM [14], [15], [23], [24], which briefly will be described below.

Let $U = (H^{in} + iI)^{-1}(H^{in} - iI)$ be the Cayley transform of $H^{in}$ at the complex point $i = \sqrt{-1}$ and $\theta$ be an arbitrary normalized generating element of $H^{in}$ in $\mathcal{H}^{in}$. The restriction of $U$ on the subspace $\mathcal{H}^{in} \ominus (U*\theta)$ we denote as $U_0$ so that $U_0 : \mathcal{H}^{in} \ominus (U*\theta) \to \mathcal{H}^{in} \ominus \{\theta\}$ and $U : U*\theta \to \theta$. Here by $\ominus$ the orthogonal difference is denoted.

Consider the linear set
$$D\left(H_0^{in}\right) \equiv \{2i(H^{in} - iI)^{-1} u, u \in \mathcal{H}^{in} \ominus \{\theta\}\}$$
in $\mathcal{H}^{in}$ and restrict the s.a. operator $H^{in}$ on the linear set $D\left(H_0^{in}\right)$:
$$H_0^{in} = H^{in} \mid D\left(H_0^{in}\right). \tag{3}$$
Here the symbol / means the operation of restriction. The operator $H_0^{in}$ defined by this formula acts in the same way as $H^{in}$. However its domain is not the whole space $\mathcal{H}^{in}$ but $D\left(H_0^{in}\right)$ only.

It is well-known that [14], [15], [23], [24]:
1) The operator $H_0^{in}$ is symmetric on $D\left(H_0^{in}\right)$ and $U_0$ is the Cayley transform of $H_0^{in}$.

2) Deficiency indices of $H_0^{in}$ corresponding to the upper and lower complex plane of the spectral parameter of this operator are equal $(1,1)$ and deficiency elements corresponding to the points $z = \pm i$ are $\theta$ and $U*\theta = \theta*$ respectively.

Let $\mathcal{N}_i^{in} = \vee \theta$ and $\mathcal{N}_{-i}^{in} = \vee \theta^*$ be the deficiency subspaces of $H_0^{in}$ corresponding to the points $z = \pm i$. Here by $\vee$ we denote the linear span strained on the corresponding vectors.

In the subspaces $\mathcal{N}^i = \mathcal{N}_i^{in} + \mathcal{N}_{-i}^{in}$ one can form a new Riesz basis $\{W^\pm\}$
$$W^+ = \frac{H^{in}}{H^{in} - i}\theta, \quad W^- = \frac{1}{H^{in} - i}\theta. \tag{4}$$

### 3.4. Spin degrees of freedom in the internal channel

Let us now introduce the spin variables in the internal channel, i.e. define the internal spin space $\mathcal{H}_s^{in} = \mathcal{H}^{in} \otimes \mathcal{H}_s$ and internal operator with spin $H_s^{in} = H^{in} \otimes I_s$. It leads to double degeneration of the energy levels $E_j$ of quantum dot and deficiency indexes of the operator $H_{0s}^{in} = H_0^{in} \otimes I_s$ become equal to $(2, 2)$. In particular, the deficiency subspace $\mathcal{N}_s^{in} \equiv \mathcal{N}^{in} \otimes \mathbb{C}^2$ becomes four-dimensional with the basis $\{W^\pm\}_\alpha, \alpha = \pm 1/2$. So the so-called boundary form $J^{in}$ of the operator $H_{0s}^{in}$ (see e.g. [14], [15], [23], [24]) can be written in the following form. For arbitrary $F, G \in \mathcal{N}_s^{in}$ it reads

$$J^{in}(G,F) \equiv \left\langle H_{0s}^{in*}G, F \right\rangle - \left\langle G, H_{0s}^{in*}F \right\rangle = \sum_{\alpha=\pm 1/2}\left[ \eta_\alpha^-(G)\overline{\eta_\alpha^+(F)} - \eta_\alpha^+(G)\overline{\eta_\alpha^-(F)} \right]. \quad (5)$$

Here $\eta_\alpha^\pm(G)$ are the so-called boundary values of the element G from deficiency subspace $\mathcal{N}_s^{in}$. In equation (5) the upper line means the complex conjugation and the star * means the operator conjugation.

The boundary values can be treated as decomposition coefficients with respect o the basis $\{W^\pm\}_\alpha$. An arbitrary element $G$ form deficiency subspace can be decomposed as

$$G = \sum_\alpha \left[ \eta_\alpha^-(G)W_\alpha^- + \eta_\alpha^+(G)W_\alpha^+ \right]. \quad (6)$$

The symplectic form $J^{in}(G,F)$ can be easily calculated using the property:

$$H_{0s}^{in*}W_\alpha^- = W_\alpha^+, \quad H_{0s}^{in*}W_\alpha^+ = -W_\alpha^-.$$

### 3.5. The restriction of the external operator $H_s^{ex}$ in the external spin space $\mathcal{H}_s^{ex}$

For every $\Psi \in D(H_s^{ex}) = W_2^2(\mathbb{R}) \otimes \mathcal{H}_s$ the operator $H_s^{ex}$ is s.a. differential one given by expression $H_s^{ex} = H^{ex} \otimes I_s$. Let us restrict $H_s^{ex}$ to the symmetric operator $H_{0s}^{ex}$ on the domain:

$$D(H_{0s}^{ex}) = \left\{ \Psi_\alpha \in W_2^2(\mathbb{R}) \otimes \mathcal{H}_s : \Psi_\alpha(0) = 0, \Psi_\alpha'(0) = 0 \right\}.$$

Then the restriction is defined as

$$H_{0s}^{ex} = H_s^{ex} \mid D(H_{0s}^{ex}).$$

The deficiency indexes of such restriction (with taking into account the spin degrees of freedom) are equal to (2, 2). It can be justified by solution of the equations

$$\left( H_{0s}^{ex} \pm i \right)\Phi = 0, \quad \Phi \in L^2(\mathbb{R}) \otimes \mathcal{H}_s.$$

At the complex points $z = \pm i$ one has two solutions belonging to $L^2(\mathbb{R}) \otimes \mathcal{H}_s$.

For every $\Phi$ from spin deficiency external subspace $\mathcal{N}_s^{ex} = (\mathcal{N}_i^{ex} + \mathcal{N}_{-i}^{ex}) \otimes \mathcal{H}_s$, $\Phi = \begin{pmatrix} \varphi_{+1/2} \\ \varphi_{-1/2} \end{pmatrix}$, one can calculate the boundary values $\xi_\alpha^\pm, \alpha = \pm 1/2$ of $\Phi$ as follows

$$\xi_\alpha^+ = \varphi_\alpha(0),$$
$$\xi_\alpha^- = \varphi_\alpha'\mid_{x=+0} - \varphi_\alpha'\mid_{x=-0}.$$

The boundary symplectic form $J^{ex}$ in external channel with spin can be obtained by integration by parts:

$$J^{ex}(\Psi,\Phi) \equiv \left\langle H_{0s}^{ex*}\Psi, \Phi \right\rangle - \left\langle \Psi, H_{0s}^{ex*}\Phi \right\rangle = \sum_{\alpha=\pm 1/2}\left[ \xi_\alpha^-(\Psi)\overline{\xi_\alpha^+(\Phi)} - \xi_\alpha^+(\Psi)\overline{\xi_\alpha^-(\Phi)} \right]. \quad (7)$$

The symplectic boundary form $J^{ex}$ is the analog of the form $J^{in}$ in the internal channel.

### 3.6. The joint extension of $H_{0s}^{ex}$ and $H_{0s}^{in}$ to the self-adjoint. operator $H_\Gamma$

Let us consider in the space $\mathcal{H} = \mathcal{H}_{0s}^{ex} \oplus \mathcal{H}_{0s}^{in} = (\mathcal{H}^{ex} \oplus \mathcal{H}^{in}) \oplus \mathcal{H}_s$ the symmetric operator

$$H_0 = H_{0s}^{ex} \oplus H_{0s}^{in} \quad (8)$$

on the domain $D(H_{0s}^{ex}) \oplus D(H_{0s}^{in})$. Here $\oplus$ stands for the orthogonal sum. The operator $H_0$ has the deficiency subspace $\mathcal{N}_s = \mathcal{N}_s^{ex} \oplus \mathcal{N}_s^{in}$. In order to extend the operator $H_0$ to the s.a. one it is enough to choose in $\mathcal{N}_s$ such linear set $\mathcal{L}_\Gamma$, which nulls the total boundary form $J = J^{ex} + J^{in}$ of the operator $H_0$, i.e.

$$J \mid \mathcal{L}_\Gamma = 0$$

It turns out that all such $\mathcal{L}_\Gamma$ can be written [23] as

$$\begin{pmatrix} -\xi_\alpha^+ \\ \eta_\alpha^- \end{pmatrix} = \Gamma \begin{pmatrix} \xi_\alpha^- \\ \eta_\alpha^+ \end{pmatrix}, \tag{9}$$

where $\Gamma = \Gamma^*$ is an arbitrary s.a. operator in $\mathcal{N}_s$. Then the operator $H_0^*$ with the boundary conditions (9) is some restriction of $H_0^*$ and this operator $H_\Gamma$ is s.a. on the domain

$$D(H_0^*) + \mathcal{L}_\Gamma = D(H_\Gamma),$$

where + stands for the direct sum.

Generally speaking the operator $\Gamma$ is an arbitrary self-adjoint operator acting in the common deficiency subspace $\mathcal{N}_s$. Due to the orthogonal structure of the channel coupling in our modeling the operator $\Gamma$ is $2 \times 2$ operator matrix

$$\Gamma = \begin{pmatrix} \gamma^{ee} & \gamma^{ei} \\ \gamma^{ie} & \gamma^{ii} \end{pmatrix} \tag{10}$$

Taking into account the degeneracy of the spectrum in both channels we choose $\gamma$ with superscripts as $2 \times 2$ numerical matrices.

### 3.7. The solution of the spectral problem for $H_\Gamma$

The spectral problem for $H_\Gamma$ reads as:

$$H_0^* \Psi = E\Psi, \Psi = \begin{pmatrix} \Psi^{ex} \\ \Psi^{in} \end{pmatrix} \in \mathcal{H} \tag{11}$$

with the boundary conditions:

$$\begin{pmatrix} -\xi_\alpha^+ \\ \eta_\alpha^- \end{pmatrix} = \Gamma \begin{pmatrix} \xi_\alpha^- \\ \eta_\alpha^+ \end{pmatrix}, \tag{12}$$

where $\xi^\pm$ are the boundary values of $\Psi^{ex}$ and $\eta^\pm$ are the boundary values of $\Psi^{in}$. It is well-known in frames of ETM that on the solution of the spectral problem written above the boundary values $\xi^+$ and $\xi^-$ as well as $\eta^+$ and $\eta^-$ are related by some additional equation. In particular,

$$\eta^- = D^{in}(E)\eta^+, \tag{13}$$

where $D^{in}(E)$ is $2 \times 2$ matrix $D_{\alpha\beta}^{in}(E)$ given by equation:

$$D_{\alpha\beta}^{in}(E) = \left\langle (I + EH_s^{in})(H_s^{in} - EI)^{-1}\theta_\alpha, \theta_\beta \right\rangle, \alpha, \beta = \pm 1/2. \tag{14}$$

The spinor structure of the internal channel chosen in the present model leads to that $D^{in}_{\alpha\beta}(E)$ turns out to be diagonal, i.e.

$$D^{in}(E) = \begin{pmatrix} D^{in}_{+1/2,+1/2} & 0 \\ 0 & D^{in}_{-1/2,-1/2} \end{pmatrix}. \tag{15}$$

Moreover, the direct calculations show that $D^{in}_{+1/2,+1/2} = D^{in}_{-1/2,-1/2}$. It is because in the simplest model of quantum dot internal structure considered here the spin-flip processes due to internal interaction are forbidden. In the spectral representation of $H^{in}$ (see equation (2)) the matrix $D^{in}(E)$ can be calculated explicitly in terms of eigenvalues $E_j$ and eigenvectors $f_j$. Namely,

$$D^{in}_{+1/2,+1/2} = D^{in}_{-1/2,-1/2} \equiv D(E) = \sum_{j=1}^{j=n} \alpha_j \frac{EE_j + 1}{E_j - E}, \tag{16}$$

where $\alpha_j$ are the mean values of eigenprojections of the Hamiltonian of quantum dot on its generating element: $|\langle \theta, f_j \rangle|^2$. Again the direct calculations show that $D(E)$ does not depend on spin variables. The reason of this fact is absence of spin-flip processes in quantum dot model developed here.

The relation (13) allows to exclude from the spectral problem (11) all ingredients of the internal channel $\mathcal{H}^{in}_s$ and reduce the whole problem to the spectral problem in $\mathcal{H}^{ex}_s$ only. Namely, using the boundary conditions (12), i.e.

$$-\xi^+ = \gamma^{ee}\xi^- + \gamma^{ei}\eta^+, \tag{17}$$
$$\eta^- = \gamma^{ie}\xi^- + \gamma^{ii}\eta^+$$

and equation (13) one obtains

$$D^{in}(E)\eta^+ = \gamma^{ie}\xi^- + \gamma^{ii}\eta^+.$$

That gives

$$\eta^+ = \left(D^{in}(E) - \gamma^{ii}\right)^{-1}\gamma^{ie}\xi^-.$$

Inserting this relation into equation (17) we obtain

$$\xi^- = \hat{\gamma}(E)\xi^+, \tag{18}$$

where by $\hat{\gamma}(E)$ we denote the energy-dependent operator

$$\hat{\gamma}(E) = -\left[\gamma^{ee} + \gamma^{ei}(D^{in}(E) - \gamma^{ii})^{-1}\gamma^{ie}\right]^{-1}. \tag{19}$$

Due to the orthogonal structure of the operator $H_0^*$:

$$H_0^* = \begin{pmatrix} H_0^{ex*} & 0 \\ 0 & H_0^{in*} \end{pmatrix} \begin{pmatrix} \Psi^{ex} \\ \Psi^{in} \end{pmatrix}$$

from equation (11) together with equation (19) we reduce the whole spectral problem in connected external and internal subspaces to the energy-dependent boundary value problem in the external channel only:

$$\left(-\frac{d^2}{dx^2}\otimes I_s - E\right)\Psi^{ex} = 0$$

$$\Psi^{ex}(+0) = \Psi^{ex}(-0)$$

$$\frac{d}{dx}\Psi^{ex}|_{x=+0} - \frac{d}{dx}\Psi^{ex}|_{x=-0} = \hat{\gamma}(E)\Psi^{ex}(0). \tag{20}$$

It should be noted that in the boundary value problem above the spin variables are already separated and it defines the coordinate parts of external channel wave functions $\Psi^{ex} = \begin{pmatrix} \psi^{ex}_{+1/2} \\ \psi^{ex}_{-1/2} \end{pmatrix}$. Formally the boundary value problem (20) can be written as operator equation

$$\left(-\frac{d^2}{dx^2}\otimes I + \delta(x)\otimes\hat{\gamma}(E) - E\right)\Psi^{ex} = 0, \tag{21}$$

where $\otimes$ stands for the tensor product. In this equation the singular interaction $V^s(E) = \delta(x)\otimes\hat{\gamma}(E)$ is an energy-dependent generalization of the original Breit-Fermi interaction used in [1].

In order to obtain the quantum dots without internal structure one has to disconnect the link between internal and external channels. To do this it is sufficient to set in the operator $\Gamma$ (see equations (10), (12)) antidiagonal matrices $\gamma^{ie}$, $\gamma^{ei}$ to be equal to zero. Than the energy-dependent operator $\hat{\gamma}(E)$ given by equation (19) is reduced to the $2\times 2$ numerical symmetric matrix $-(\gamma^{ee})^{-1}$. This will correspond to the spin filter device constructed from the quantum dots without internal structure.

## 4. Scattering on periodic array of *N* quantum dots with internal structure

In this section we generalize the results obtained above in case of *N* quantum dots with arbitrary internal structure. First of all we need to consider problem (20) for this case. After separation of spin variables in equation $H^{(N)}\Psi^{ex(N)} = E\Psi^{ex(N)}$ one obtains for the coordinate part $\Psi^{ex(N)} = \begin{pmatrix} \psi^{ex(N)}_{+1/2} \\ \psi^{ex(N)}_{-1/2} \end{pmatrix}$ of the total wave function the following boundary value problem in external channel

$$\left(-\frac{d^2}{dx^2}\otimes I_s - E\right)\Psi^{ex(N)} = 0 \tag{22}$$

$$\Psi^{ex(N)}(y_l+0) = \Psi^{ex(N)}(y_l-0),$$

$$\frac{d}{dx}\Psi^{ex(N)}|_{x=y_l+0} - \frac{d}{dx}\Psi^{ex(N)}|_{x=y_l-0} = \hat{\gamma}(E)\Psi^{ex(N)}(y_l), l=1,2,...,N. \tag{23}$$

Note that here instead of equation $H^{(N)}\Psi^{ex(N)} = E\Psi^{ex(N)}$ it is more convenient to use expression $H^{(N)}\Psi^{ex(N)} = k^2\Psi^{ex(N)}$, where $k^2 = E2m^*/\hbar^2$. Moreover we choose the numerical matrices $\gamma^{ie}$, $\gamma^{ei}$, $\gamma^{ee}$, $\gamma^{ii}$ from (19) so that the antidiagonal elements in them are equal to zero. Hence the interaction matrix $\hat{\gamma}(E)$ takes the form

$$\hat{\gamma}(E) = \begin{pmatrix} \gamma_{11}(E) & 0 \\ 0 & \gamma_{00}(E) \end{pmatrix}. \tag{24}$$

As a result in the proposed model in the absence of relativistic effects the total spin $S$ conserves and the triplet-singlet and singlet-triplet transitions are forbidden. So that the spin-orbital interaction is not taken into account here.

Let us calculate the transmission coefficients in the singlet and triplet channels. Obviously, solution of equation (22) between the points $y_l$ and $y_{l+1}$ has the form

$$\psi^{ex(N)}(x) = e^{ikx}\sum_{\beta} A_{\alpha\beta}^{ex(l)}(k)\upsilon_{\beta} + e^{-ikx}\sum_{\beta} B_{\alpha\beta}^{ex(l)}(k)\upsilon_{\beta}. \qquad (25)$$

Here the subscripts $\alpha, \beta$ can take the values $\pm$ and two-component vectors $\upsilon_{+}(j) = \begin{pmatrix} 1 \\ 0 \end{pmatrix}$, $\upsilon_{-}(j) = \begin{pmatrix} 0 \\ 1 \end{pmatrix}$. The coefficients $A_{\alpha\beta}^{ex(l)}(k)$ and $B_{\alpha\beta}^{ex(l)}(k)$ are to be determined due to matching conditions (23). Let us denote by $A_{\alpha\beta}^{ex(0)}(k)$ and $B_{\alpha\beta}^{ex(0)}(k)$ the corresponding coefficients of the function $\psi^{ex(N)}(x)$ when $x < y_1$ and by $A_{\alpha\beta}^{ex(N)}(k)$, $B_{\alpha\beta}^{ex(N)}(k)$ the coefficients when $x > y_N$. If one sets $A_{++}^{ex(0)}(k) = A_{--}^{ex(0)}(k) = 1$, $A_{+-}^{ex(0)}(k) = A_{-+}^{ex(0)}(k) = 0$ (when incident wave has the unit amplitude) and $B_{\alpha\beta}^{ex(N)}(k) = 0$ (there is only transmitted wave on the right of the array of quantum dots) than we have 4 transmission coefficients: $T_{11}^{ex(N)}(k) = A_{11}^{ex(N)}(k)$, $T_{00}^{ex(N)}(k) = A_{--}^{ex(N)}(k)$, $T_{10}^{ex(N)}(k) = A_{+-}^{ex(N)}(k)$ and $T_{01}^{ex(N)}(k) = A_{-+}^{ex(N)}(k)$. Two last coefficients can be interpreted as spurious solutions describing the triplet-singlet and singlet-triplet transitions. As it has been mentioned above in the present model the total spin $S$ conserves and spin-flip processes are forbidden. Thus the transmission coefficients $T_{00}^{ex(N)}, T_{11}^{ex(N)}$ are calculated in terms of elements of $4 \times 4$ matrix $\hat{T} = t_{ij}, i,j = 1,2,3,4$ ($i$ is the line number, $j$ is the column number) which is the product of transition matrices through dots and wires connected dots. It reads $\hat{T} = Q^{N-1}Q_1$, where

$$Q = \frac{1}{2ik}\begin{pmatrix} \gamma_{11}(E)+2ik & 0 & e^{-2idk}\gamma_{11}(E) & 0 \\ 0 & \gamma_{00}(E)+2ik & 0 & e^{-2idk}\gamma_{00}(E) \\ -\gamma_{11}(E) & 0 & e^{-2idk}(2ik-\gamma_{11}(E)) & 0 \\ 0 & -\gamma_{00}(E) & 0 & e^{-2idk}(2ik-\gamma_{00}(E)) \end{pmatrix}. \qquad (26)$$

The matrix $Q_1$ differs from $Q$ by the absence of multiplier $e^{-2idk}$ in its matrix elements. Here $d$ is the length of the quantum wire connecting two neighbor quantum dots. The transmission coefficients in external channel have the form $T_{11}^{ex(N)} = t_{11} + t_{13}e^{-2iy_1k}R_{11}^{ex(N)}$, $T_{00}^{ex(N)} = t_{22} + t_{24}e^{-2iy_1k}R_{00}^{ex(N)}$. The reflection coefficients $R_{11}^{ex(N)}, R_{00}^{ex(N)}$ are given by equations $R_{11}^{ex(N)} = e^{2iy_1k}M^{-1}(t_{41}t_{34} - t_{31}t_{44})$, $R_{00}^{ex(N)} = e^{2iy_1k}M^{-1}(t_{43}t_{32} - t_{33}t_{42})$, where $M = t_{33}t_{44} - t_{34}t_{43}$. In addition the relations which imply the unitary property of scattering have been checked, i.e. $|T_{jj}^{ex(N)}|^2 + |R_{jj}^{ex(N)}|^2 = 1, (j=0,1)$.

## 5. Results and discussions

In this section we present the results of numerical calculations of the operational regime characteristics for the spin gun and discuss the proper choice of the model parameters as well as the choice of semiconductor materials for QW and QD. We use the applied bias $V$: $eV = \mu_L - \mu_R$ for tuning the spin polarization effects in the device. Since we use the assumption of mono-mode ballistic electron propagation in QW connected the quantum dots we have to mention that if one use the bias applied to the device the following physical limitations have to be taken into account. In mono-mode ballistic regime of spin gun operation the resistance will be at least $13\,\text{k}\Omega\,(\pi\hbar/e^2)$ for $GaAs$ QW [25]. Hence in order to reach decent current level the fairly large bias has to be applied. At the same time when the relatively large bias is applied to semiconductor QW any ballistic electron will gain enough energy to emit an optic phonon. For example, taking $36\,\text{meV}$ as upper limit (for $GaAs$ QW) of applied bias we are limited to currents of less than $2.7\,\mu\text{A}$ [25]. In order to overcome these limitations one can use other narrow-gap semiconductors listed in table 1. In particular, table 1 shows that the best choice is $InSb$ semiconductor.

In order to calculate the conductance of the device $G$ we exploit the two-channel Landauer formula at zero-temperature limit:

$$G = \frac{e^2}{\pi\hbar}\left[\left|T_{00}^{ex(N)}(E_F + eV)\right|^2 + \left|T_{11}^{ex(N)}(E_F + eV)\right|^2\right].$$

For unpolarized incident beam of electrons the scattering by QD array in the triplet and singlet channels leads to different transmission coefficients $T_{11}^{ex(N)}$ and $T_{00}^{ex(N)}$. Following [26] we shall use the polarization efficiency

$$P = \frac{\left|T_{11}^{ex(N)}\right|^2 - \left|T_{00}^{ex(N)}\right|^2}{\left|T_{11}^{ex(N)}\right|^2 + \left|T_{00}^{ex(N)}\right|^2}$$

which determines the difference of transmission probabilities through the spin gun for the triplet and singlet spin states.

At a given wave number $k$ the conductance $G$ and polarization efficiency $P$ of the device both depend on the model parameters: the distance $d$, the coupling constants with external channel $\gamma_{11}(E)$, $\gamma_{00}(E)$, the internal energy levels $E_j$ (16) of quantum dots and the number $N$ of QD. They also both depend on the choice of semiconductor materials from which the QD and QW are fabricated, i.e. effective mass $m^*$, the Fermi energy $E_F$ and the mean size $r_0$ of quantum dots. Varying the bias applied one can calculate the dependence of $G$ and $P$ on $k = \left[2m^*(E_F + eV)/\hbar^2\right]^{1/2}$ at fixed model parameters and chosen $m^*$, $E_F$, $E_j$ and $r_0$. Let us mention that from spin analysis in [16] in case of quantum dots without internal structure one has $\gamma_{00} = -3\gamma_{11}$ in (24). That is why the relations between the diagonal elements in matrices $\gamma^{ee}$, $\gamma^{ei}$, $\gamma^{ii}$, $\gamma^{ie}$ in (19) are chosen the same way.

In the figures 3 - 6 we show the conductance $G$ and polarization efficiency $P$ versus dimensionless momentum $\hat{k} = kd/\pi$ of incident electrons in the case of $N$ quantum dots. The values of $N$ and interdot distance $d$ are given in each figure capture. By the dashed line the Fermi level $E_F = 35\,\mathrm{meV}$ recalculated into the corresponding value of dimensionless momentum $\hat{k}_F = d/\pi \left[ 2m^* E_F / \hbar^2 \right]^{1/2}$ is indicated for $InSb$ semiconductor for QW and QD.

The general behavior of the conductance $G$ and polarization efficiency $P$ for the wide range of parameters variation is the following. On the $\hat{k}$ - axis with the growth of the number $N$ of quantum dots there arise the so-called "working windows", i.e. such intervals in which $P$ and $G$ are sufficiently high simultaneously. Due to the interference processes in a QD array at $N \sim 10 \div 15$ the values of $P$ and $G$ on the working windows approach $P \approx 100\%$ and $G = e^2 / \pi \hbar$ which is one half of its upper limit. It means that at $\hat{k}$ lying in the working windows the spin gun really produces the fully polarized spin beam and conducts at fairly high level. All the figures show that for different sets of the model parameters one can achieve that the Fermi level would locate inside one of the working windows in which $P$ and $G$ are sufficiently high simultaneously. The working windows extend on the right of $\hat{k}_F$ up to the value of $\hat{k} = d/\pi \left[ 2m^* (E_F + eV) / \hbar^2 \right]^{1/2}$ which corresponds to the bias applied.

Finally, let us remind that in our model all the quantum dots forming the array are exactly identical. It is obvious that no qualitative changes will occur in the spin gun operation regime if the dots are only approximately equivalent or the interdot distances are only approximately the same.

We have demonstrated that the proposed mathematical model of spin-spin interaction in the effective Hamiltonian for the spin gun yields a high spin polarization of electrons transmitted through the device. The effect could be employed for manufacturing spin filters of nanosize on the base of narrow-gap semiconductors.

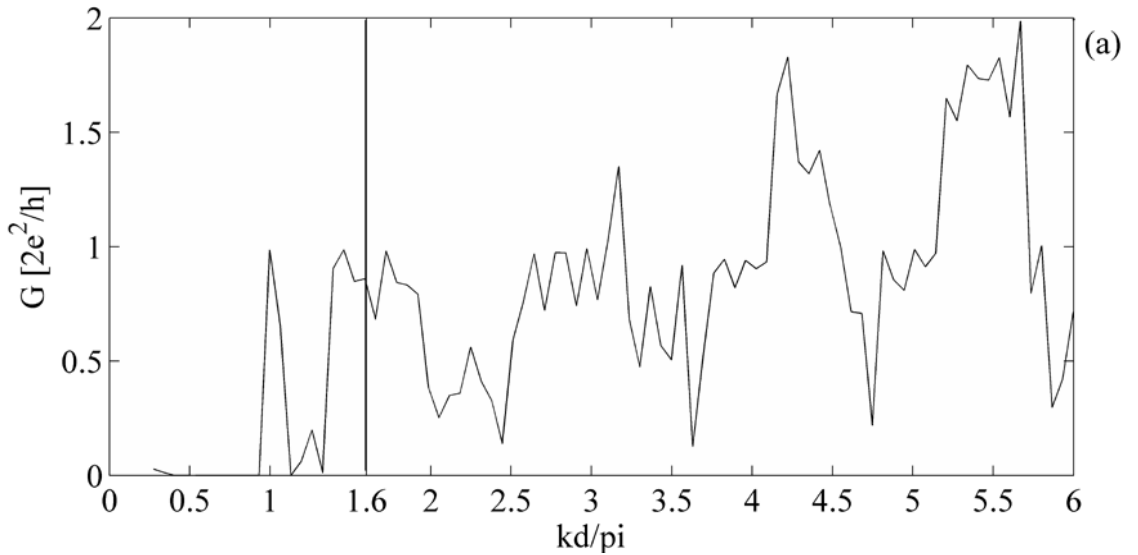

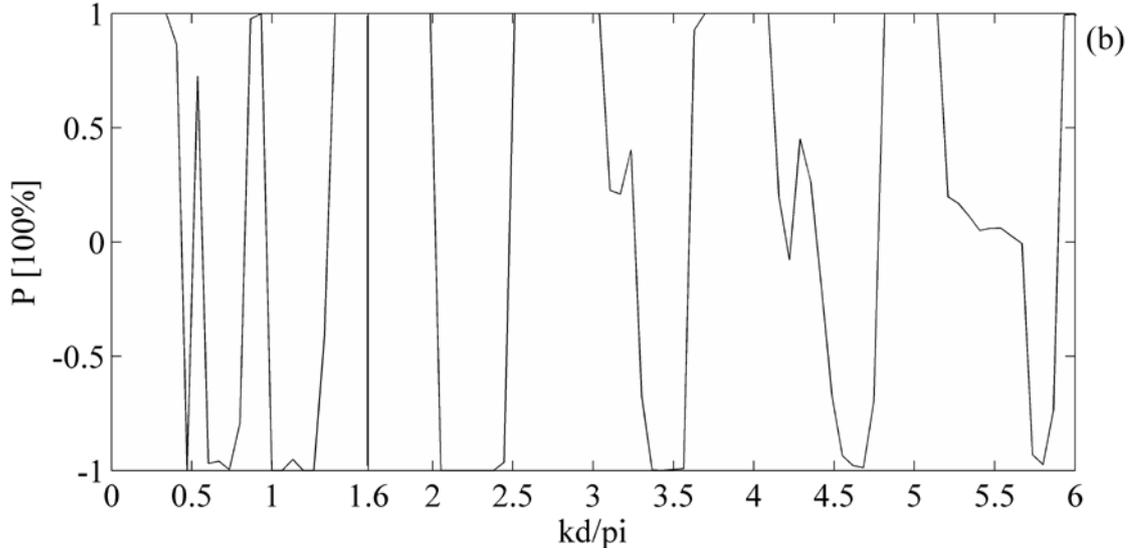

Figure 3. The conductance $G$ (a) and polarization efficiency $P$ (b) versus dimensionless momentum $\hat{k} = kd/\pi$ of incident electrons in the case of 10 quantum dots with internal structure. By the solid vertical line the Fermi level $E_F = 35\,\mathrm{meV}$ in $InSb$ is indicated. The interdot distance $d = 45\,\mathrm{nm}$, the internal energy levels $E_1 = 0.1\,\mathrm{meV}$, $E_2 = 0.9\,\mathrm{meV}$, the coefficients $\alpha_1 = 0.5$, $\alpha_2 = 0.9$.

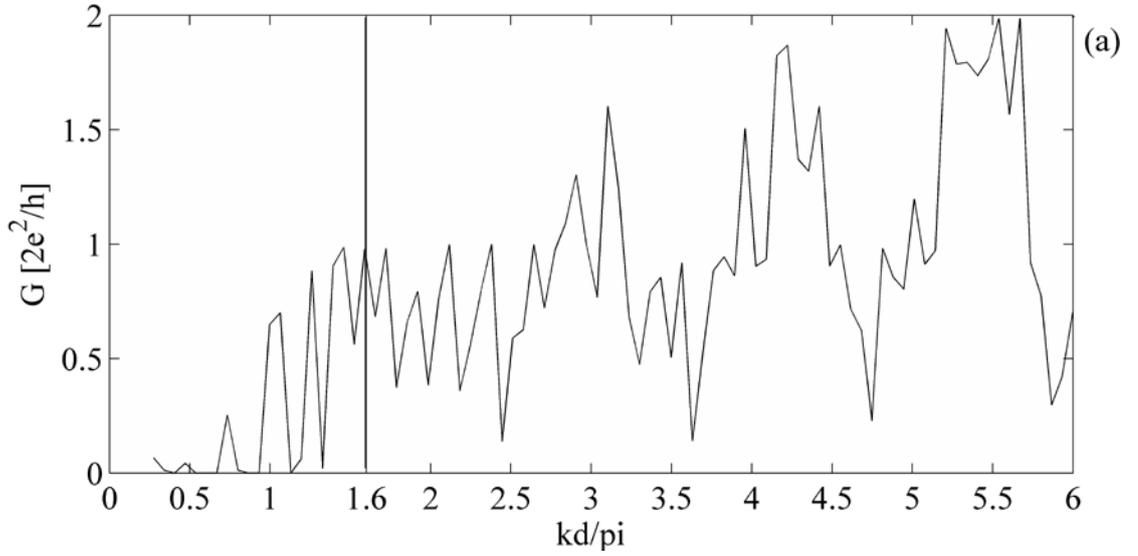

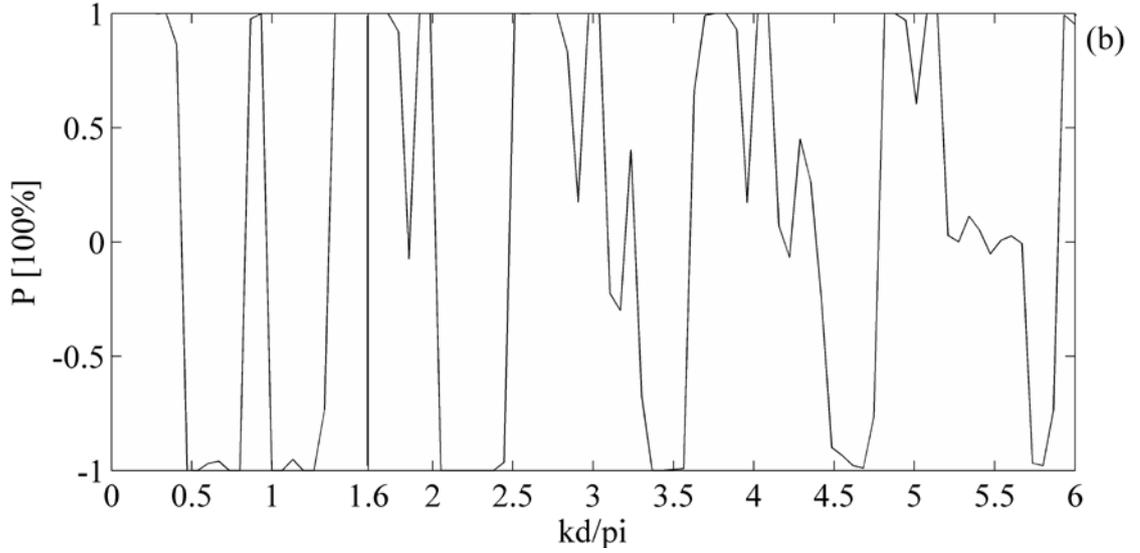

**Figure 4.** The conductance $G$ (a) and polarization efficiency $P$ (b) versus dimensionless momentum $\hat{k} = kd/\pi$ of incident electrons in the case of 10 quantum dots with internal structure. By the solid vertical line the Fermi level $E_F = 35\,\text{meV}$ in *InSb* is indicated. The interdot distance $d = 45\,\text{nm}$, the internal energy levels $E_1 = 0.5\,\text{meV}$, $E_2 = 0.6\,\text{meV}$, $E_3 = 0.7\,\text{meV}$, $E_4 = 0.8\,\text{meV}$, $E_5 = 0.9\,\text{meV}$, the coefficients $\alpha_1 = 0.5$, $\alpha_2 = 0.6$, $\alpha_3 = 0.7$, $\alpha_4 = 0.8$, $\alpha_5 = 0.9$.

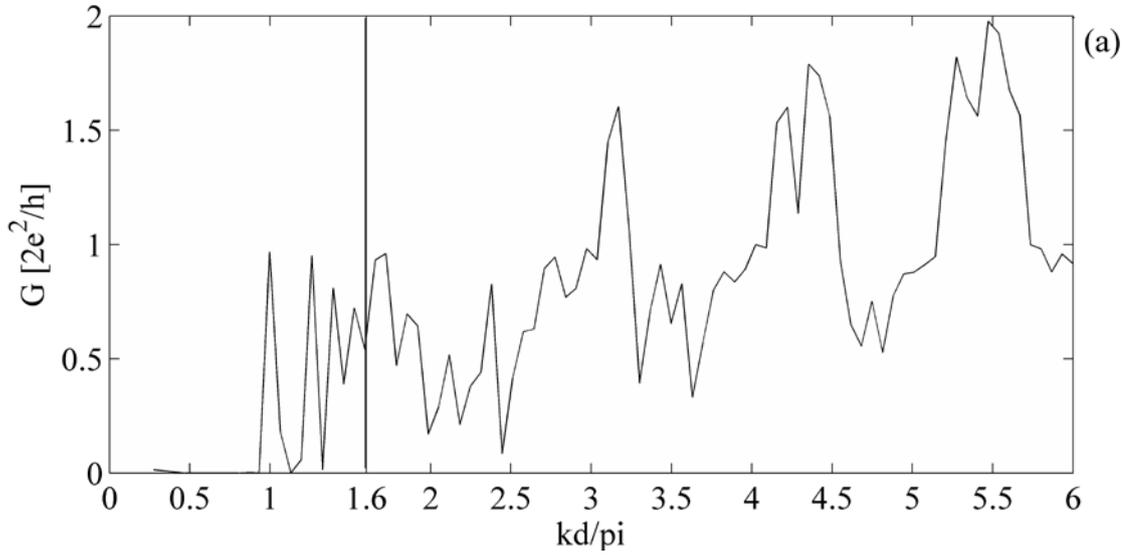

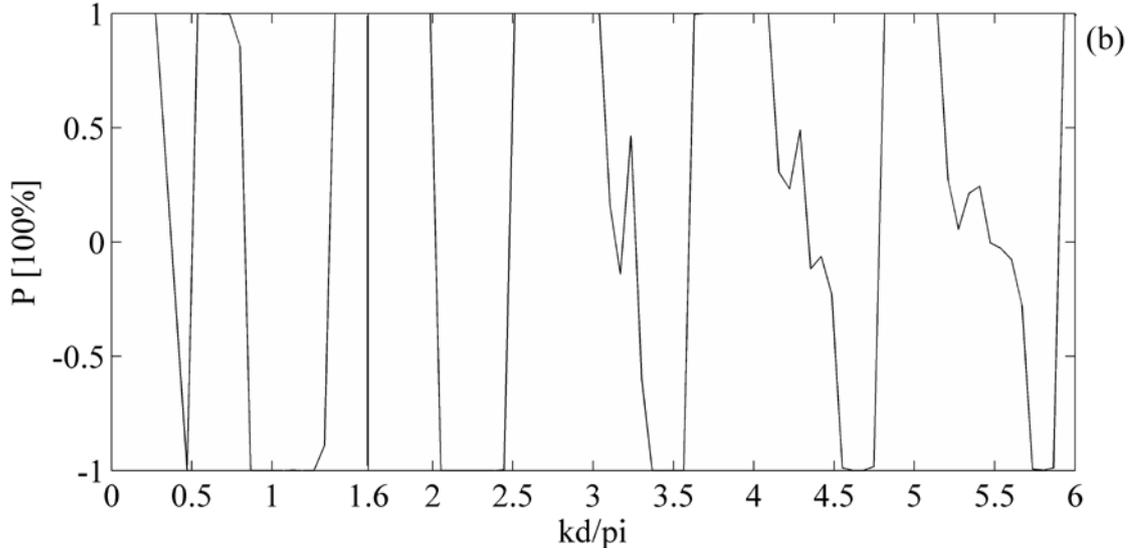

**Figure 5.** The conductance $G$ (a) and polarization efficiency $P$ (b) versus dimensionless momentum $\hat{k} = kd/\pi$ of incident electrons in the case of 15 quantum dots with internal structure. By the solid vertical line the Fermi level $E_F = 35\,\text{meV}$ in $InSb$ is indicated. The interdot distance $d = 45\,\text{nm}$, the internal energy levels $E_1 = 0.1\,\text{meV}$, $E_2 = 0.9\,\text{meV}$, the coefficients $\alpha_1 = 0.5$, $\alpha_2 = 0.9$.

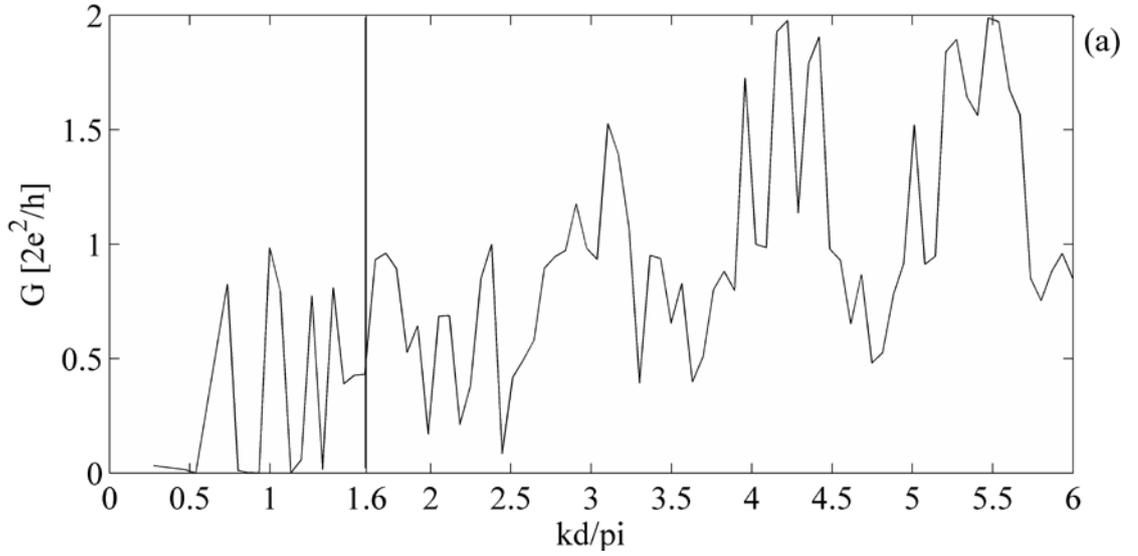

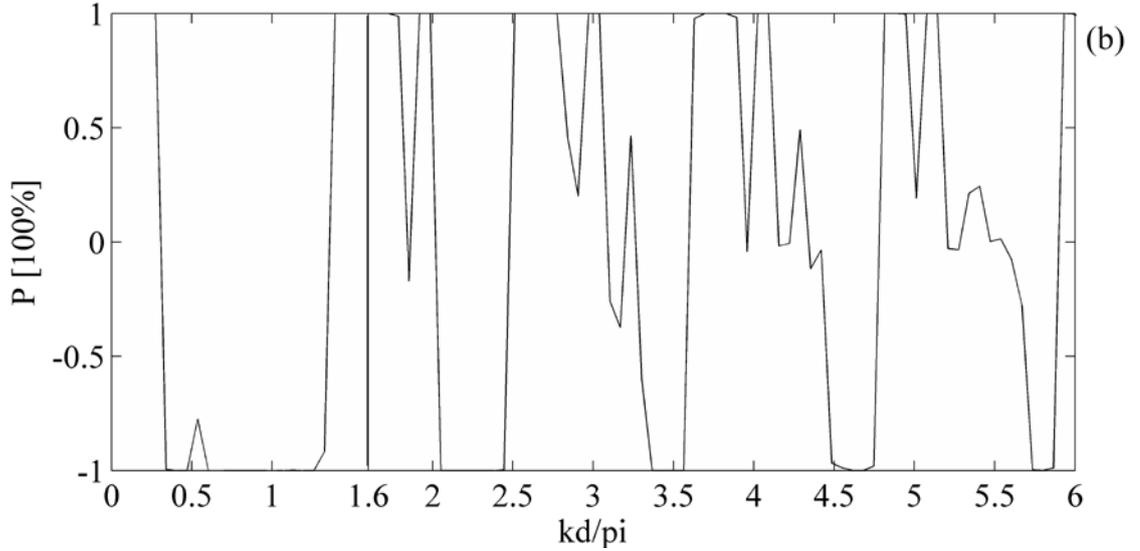

**Figure 6.** The conductance $G$ (a) and polarization efficiency $P$ (b) versus dimensionless momentum $\hat{k} = kd/\pi$ of incident electrons in the case of 15 quantum dots with internal structure. By the solid vertical line the Fermi level $E_F = 35\,\mathrm{meV}$ in $InSb$ is indicated. The interdot distance $d = 45\,\mathrm{nm}$, the internal energy levels $E_1 = 0.5\,\mathrm{meV}$, $E_2 = 0.6\,\mathrm{meV}$, $E_3 = 0.7\,\mathrm{meV}$, $E_4 = 0.8\,\mathrm{meV}$, $E_5 = 0.9\,\mathrm{meV}$, the coefficients $\alpha_1 = 0.5$, $\alpha_2 = 0.6$, $\alpha_3 = 0.7$, $\alpha_4 = 0.8$, $\alpha_5 = 0.9$.